\documentclass[journal,10pt]{IEEEtran}
\usepackage{cite}      
\usepackage{graphicx}  
\usepackage{url}
\usepackage{balance}
\usepackage{times}
\usepackage{amsmath}   
\usepackage{pdflscape}
\usepackage[normalem]{ulem}
\usepackage{mathtools}
\usepackage{nicefrac}
\usepackage[none]{hyphenat}
\usepackage{array}
\usepackage{fontenc}
\usepackage{color}
\usepackage{etoolbox}
\usepackage{bm}
\usepackage{comment}
\usepackage{caption}
\usepackage{subcaption}
\usepackage{amsmath,amsthm,amssymb, amsfonts}
\usepackage{mathtools}
\usepackage{multirow}
\usepackage{wrapfig}
\usepackage{tikz}
\usepackage{tabularx}
\usepackage{booktabs}

\usetikzlibrary{arrows.meta, positioning, shapes.geometric}

\usepackage{nomencl}
\usepackage{algorithm}
\usepackage{algpseudocode}
\usepackage{accents}
\usepackage{acronym}

\newlength{\dhatheight}

\usepackage{dblfloatfix}
\newcolumntype{P}[1]{>{\centering\arraybackslash}p{#1}}
\newcolumntype{M}[1]{>{\centering\arraybackslash}m{#1}}
\usepackage{verbatim} 
\usepackage{float}
\usepackage{placeins}
\usepackage{comment}
\usepackage{dblfloatfix}
\usepackage{pifont}
\usepackage{xparse}
\makeatletter
\acrodef{3D}{Three Dimensional}
\acrodef{UAV}{Unmanned Aerial Vehicles}
\acrodef{ISAC}{Integrated Sensing and Communications}
\acrodef{SAR}{Synthetic Aperture Radar}
\acrodef{MOCOMP}{Motion Compensation}
\acrodef{RCS}{Radar Cross Section}
\makeatother
\begin{document}

\setlength{\textfloatsep}{0pt}
\setlength{\intextsep}{10pt plus 2pt minus 2pt}

\title{Bistatic UAV-Based \ac{3D} SAR Imaging Using Integrated Sensing and Communications Transceivers}

\author{Akanksha Sneh,~\IEEEmembership{ Member~IEEE}, Krishna Ayyagari$^{\ast}$, Vivaswan Nawani$^{\ast}$, Shobha Sundar Ram,~\IEEEmembership{Senior Member~IEEE}.
\thanks{$^{\ast}$ Krishna Ayyagari and Vivaswan Nawani are joint second authors.}
\thanks{The first three authors are with the Indraprastha Institute of Information Technology, Delhi, New Delhi 110020 India. E-mail: \{akankshas, krishna21158, vivaswan21217\}@iiitd.ac.in. Last author is with University College London, United Kingdom, Email: \{s.ram@ucl.ac.uk\}}}

\maketitle

 \begin{abstract}
Integrated sensing and communication (ISAC) frameworks enable autonomous systems to jointly perform radar sensing and wireless communication using common spectral and hardware resources. Prior art on ISAC-enabled \ac{UAV} for \ac{SAR} focused on monostatic configurations for two-dimensional target imaging. In this work, we propose a bistatic UAV tomographic SAR framework in which the IEEE~802.11ad waveform, RF hardware, and spectral resources are jointly utilized for communication and radar sensing for three-dimensional (3D) localization of complex targets. Specifically, we demonstrate how communication packets transmitted from a master ISAC transceiver UAV to a passive receiver UAV are exploited for timing and frequency synchronization, motion compensation, and 3D localization of targets, without requiring auxiliary infrastructure on the passive receiver for these tasks. 
We evaluate the proposed framework for localizing automotive targets over elevated terrain, with Monte Carlo simulations modeling synchronization and platform motion errors. The results demonstrate improved localization accuracy compared with conventional InSAR-based localization. The interplay between radar and communication performances are further evaluated using diverse modulation schemes.

\end{abstract}


\section{Introduction}
\acp{UAV} equipped with SAR are increasingly used for remote sensing, surveillance, and situational awareness  \cite{bekar2021low,alotaibi2019lsar,fernandez2018synthetic,stecz2020uav}. These systems combine platform agility, fast deployment, and low cost with SAR benefits of all-weather, day-night imaging, and wide-area coverage.
UAV-SAR is particularly useful when traditional space-borne SAR is limited by long satellite revisit times and when manned aircraft are difficult, unsafe, or expensive to operate. Furthermore, recent advances in compact millimeter wave (mmWave) transceivers, lightweight antennas, onboard processing, and autonomous operation enable the integration of high-performance mmWave SAR payloads on small-to-medium \acp{UAV}. The large available bandwidth at mmWave frequencies provides fine range resolution, enabling high-resolution imaging \cite{wang2009waveform}.\\
\indent Monostatic \ac{UAV} SAR systems, in which the transmitter and receiver are on the same platform, are the most popular SAR systems and are easiest to deploy in terms of synchronization  \cite{cumming2005digital, bekar2021low}. However, they are limited to facilitating two-dimensional imaging along range and cross-range. Interferometric \cite{bamler1998synthetic, lahmeri2025uav}, bistatic \cite{willis2005bistatic, soumekh1991bistatic}, and multistatic  \cite{liang2022phase, kanz2025coherent} configurations exploit additional spatial, Doppler, and scattering diversity to enhance radar detection metrics and can be configured for three-dimensional (3D) scene reconstruction. The theoretical foundations of bistatic image formation through inversion was studied in \cite{soumekh1991bistatic}.
Other developments in \ac{3D} SAR imaging include interferometric SAR (InSAR) to map heights of structures \cite{bamler1998synthetic}, differential InSAR to monitor deformations \cite{wang2020ground} on the Earth's surface, and SAR tomography (TomoSAR) to resolve multiple scatterers along the elevation axis, which is also the focus of our work.
Early demonstrations of TomoSAR used multiple airborne baselines to synthesize a second aperture across tracks for optimally reconstructing elevation profiles \cite{reigber2002first}. Later developments used multipass spaceborne designs, capable of interferometry, focusing on long-term repeat-pass data to allow \ac{3D} focusing \cite{fornaro2005three}.  
The authors in \cite{lombardini2013temporal} introduced differential TomoSAR schemes based on multidimensional spectral estimation and multifaceted signal classification algorithms to address temporal decorrelation.
More recent methods used holographic SAR tomography with circular airborne trajectories and iterative adaptive reconstruction to provide super-resolution elevation reconstruction \cite{feng2020holographic}. While these advances have significantly improved the \ac{3D} imaging capability of SAR systems, their practical realization on distributed UAV platforms introduces several implementation challenges. Among these, maintaining synchronization between spatially separated platforms is one of the most challenging requirements for coherent SAR image formation.

Existing synchronization approaches for bistatic \ac{UAV} SAR focus on hardware solutions, involving specialized microwave links, reference beacons, pilot exchanges, or precise timing sources based on global navigation satellite systems or atomic oscillators on both UAVs \cite{chen2020implementation, liang2022phase, rodrigues2025experimental}. Alternatively, to reduce the hardware load, signal-processing-based algorithms that estimate and compensate for residual timing and phase errors directly from the received echoes, have been proposed. This approach is, however, effective only with high signal-to-noise ratio (SNR) and stable geometries  \cite{wang2025coherence, gungor2022compressive}.
Recently, a third approach has been investigated, in which \acp{UAV} are augmented with communication links to address synchronization and data-handling challenges in a distributed architecture. In \cite{lahmeri2025uav}, \ac{UAV}-to-ground and \ac{UAV}-to-\ac{UAV} connections facilitate real-time data transfer and sensing-aware formation control for bistatic InSAR \cite{lahmeri2025uav}. Coherent multistatic \ac{UAV} SAR architectures have also been explored in \cite{kanz2025coherent}, in which a direct line-of-sight sidelink was used as a phase reference for hardware synchronization. 

But these approaches do not scale to tomographic or \ac{3D} localization. Bistatic TomoSAR has more recently been achieved by repurposing an uplink communication signal to reconstruct tomographic elevation images \cite{yin2025tomographic}. The authors of \cite{hu2023imaging} used orthogonal frequency-division multiplexing (OFDM) pilots over line-of-sight links to provide timing and phase references for bistatic SAR processing at a ground-based station. Specialized two-way synchronization chains with pulsed carrier exchanges have also enabled high-precision bistatic InSAR in-flight experiments \cite{zhu2024unmanned}. These studies have focused on the coexistence of radar and communication systems, where communication waveforms are used for synchronization, while separate conventional radar waveforms are used for SAR imaging. Instead of using auxiliary communication platforms as discussed above, it is advantageous to consider ISAC transceivers to manage synchronization and interference between communications and radar. 

Recent studies have explored waveform co-design strategies to enable joint wireless communication and high-resolution SAR imaging, in monostatic configurations or with centralized or ground-based infrastructure \cite{chen2020implementation,wang2019first}. As a result, they do not address the challenges of distributed bistatic synchronization and motion compensation in \ac{UAV}-based tomographic SAR scenarios. 
In addition, most prior work is limited to two-dimensional imaging or simplified point-target models and does not address three-dimensional localization of complex, distributed targets within an ISAC framework. Furthermore, many studies assume stationary or ground-based receivers and simplified motion models, which limit their applicability to realistic \ac{UAV} deployments involving mobile platforms, elevated terrain, and dynamic flight trajectories. 

In practical bistatic SAR systems, platform and target motion cause significant image displacement, defocusing, and localization errors whose severity depends on the bistatic geometry and target kinematics \cite{duman2015moving,zhang2020novel}. Hence, when multiple UAVs are deployed, each has to be mounted with auxiliary navigation sensors for motion compensation.
In our proposed work, we overcome the above limitations by introducing a unified ISAC-based bistatic TomoSAR framework that leverages a single waveform, shared hardware, and common spectral resources for both communication-assisted synchronization and motion compensation and radar sensing, across multiple UAVs, while enabling accurate \ac{3D} localization under realistic operating conditions. The key advantage of our proposed scheme is that the first UAV is equipped with an ISAC transceiver and navigational sensors. In contrast, the second UAV is a simple passive receiver with no additional hardware for synchronization or motion compensation. The same architecture can, therefore, be extended to multistatic UAV configurations with little or no increase in hardware complexity. \\
\indent The key contributions of the proposed work, at the architecture, algorithmic, and system levels, are summarized as follows: (1) \textit{Architecture Innovations:} A unified bistatic TomoSAR architecture is developed on a mmWave 802.11ad ISAC platform, with shared RF hardware and spectral resources for communication and SAR. Here, the ISAC transceiver on a master UAV transmits communication packets to facilitate synchronization and motion compensation at a passive receiver UAV. This removes the need for auxiliary synchronization and navigational sensors on the slave UAV while enabling coherent bistatic operation. (2) \textit{Algorithmic Innovations:} A geometric \ac{3D} localization approach, exploiting the communication link between the UAVs, based on sphere-plane intersection, is proposed to accurately estimate scatterer positions in bistatic tomographic geometry, enabling reliable reconstruction of complex and distributed targets. We benchmark the approach against the popular frequency-modulated continuous-wave (FMCW) radar signal using InSAR signal-processing methods. The results show that our proposed method outperforms benchmark configurations across all three spatial dimensions (elevation, along-track, and across-track). For example, under motion-compensated conditions and a high platform velocity standard deviation (8~m/s), the proposed method achieves an 84.3\% reduction in the root mean square error (RMSE) for elevation when compared to the conventional FMCW-InSAR baseline
(3) \textit{High Fidelity Simulation Framework:}
To validate the proposed architecture and algorithm, we have used a high-fidelity simulation model that incorporates realistic extended targets on uneven terrain through digital elevation models (DEM), with shadowing. Further, the UAV trajectory model incorporates translational and rotational platform motion errors along with sensor errors. 
(4) \textit{Scientific insights on interplay between sensing and communication:} The use of an ISAC waveform instead of separate radar and communication systems results in a unique interdependence/interplay between the performance metrics of both functionalities. Specifically, we study how communication modulation schemes affect the radar localization metrics. \\
\indent The paper is organized as follows. In Section~\ref{sec:SysArchitecture}, we present a comprehensive overview of the proposed ISAC-based bistatic \acp{UAV} system model. Then, we discuss the signal model and processing, along with the novel 3D localization algorithm, in Section~\ref{sec:signalmodel}. In Section~\ref{sec:simsetup}, we present the high-fidelity simulation setup for evaluating the proposed framework's performance, followed by the timing analysis. Next, we present the ISAC performance metrics for different scenarios in Section~\ref{sec:results}, followed by an in-depth discussion of the novel scientific findings and a discussion on future works in Section.\ref{sec:Conclusion}.\\
\indent\emph{Notation:} Continuous and discrete time signals are represented by $(\cdot)$ and $[\cdot]$ respectively. Scalar variables, vectors, and matrices are represented by $x$, $\mathbf{x}$, and $\mathbf{X}$, respectively. The superscript $*$ corresponds to complex conjugate signal. 
\section{Proposed System Model}
\label{sec:SysArchitecture}
\begin{figure}[htbp]
    \centering
    \includegraphics[scale=0.38]{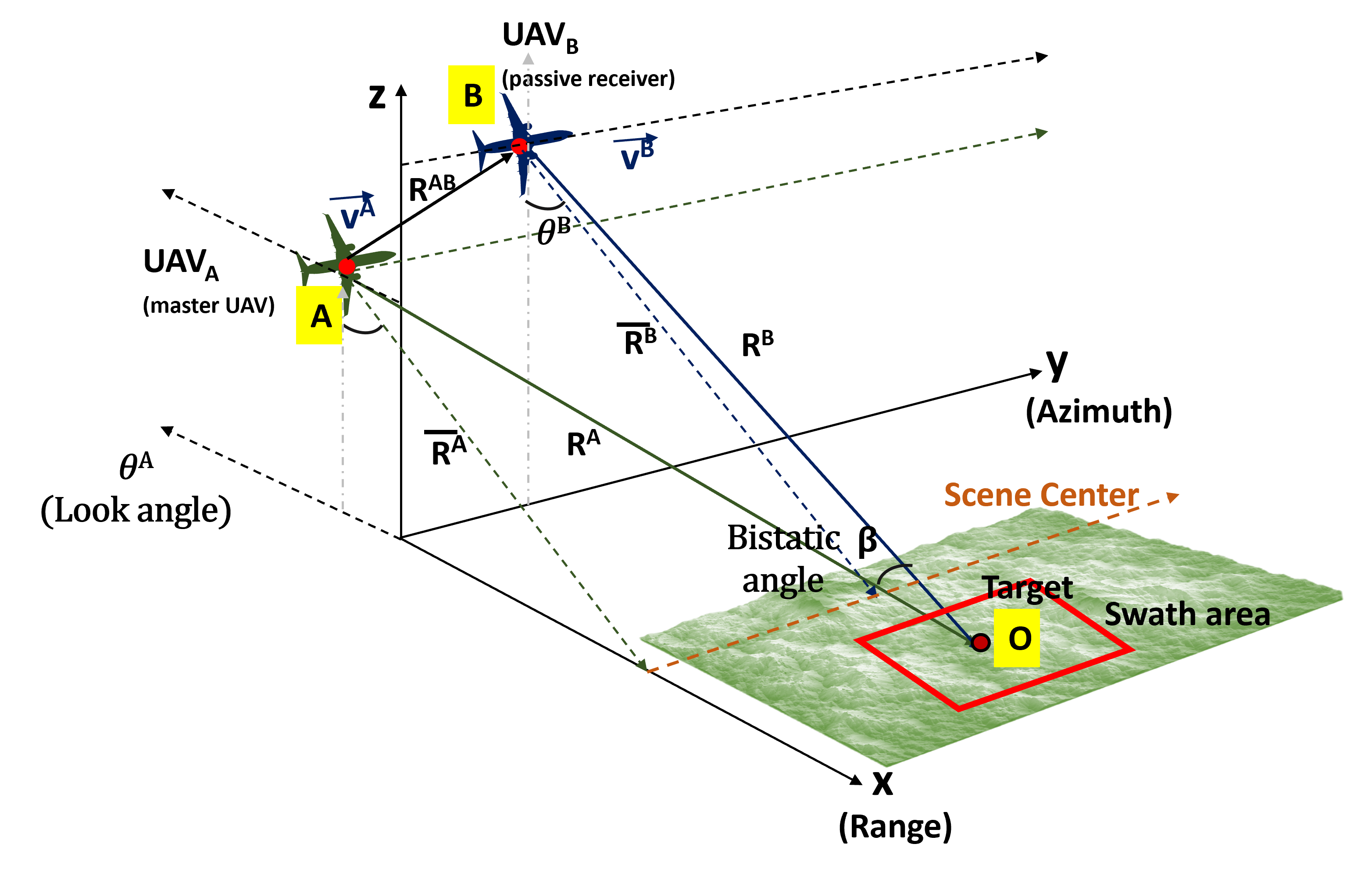}
   \vspace{-2mm}
    \caption{System model of proposed bistatic UAV-ISAC framework.}
    \label{fig:system_model}
\end{figure}

We illustrate the proposed architecture in Fig.~\ref{fig:system_model}.
A right-handed Cartesian coordinate system represents the system model, where the $x$, $y$, and $z$-axes denote the ground-range (cross-track) direction, the azimuth (along-track) direction, and the altitude, respectively. We consider two \ac{UAV} platforms, $UAV_A$, equipped with a master ISAC transceiver and navigational sensors, while $UAV_B$ only has passive receivers. Both UAVs operate over a common ground swath area and exploit a shared ISAC waveform to jointly enable synchronization through direct communication and bistatic radar sensing of ground-based targets. They fly at low altitudes under which the flat-Earth assumption holds and follow linear trajectories parallel to the azimuth direction. They are configured in a stripmap SAR mode, where the radar antenna beam of each \ac{UAV} is fixed relative to the platform and continuously illuminates the ground swath during flight.
\begin{figure*}[htbp]
    \centering
    \includegraphics[scale=0.6]{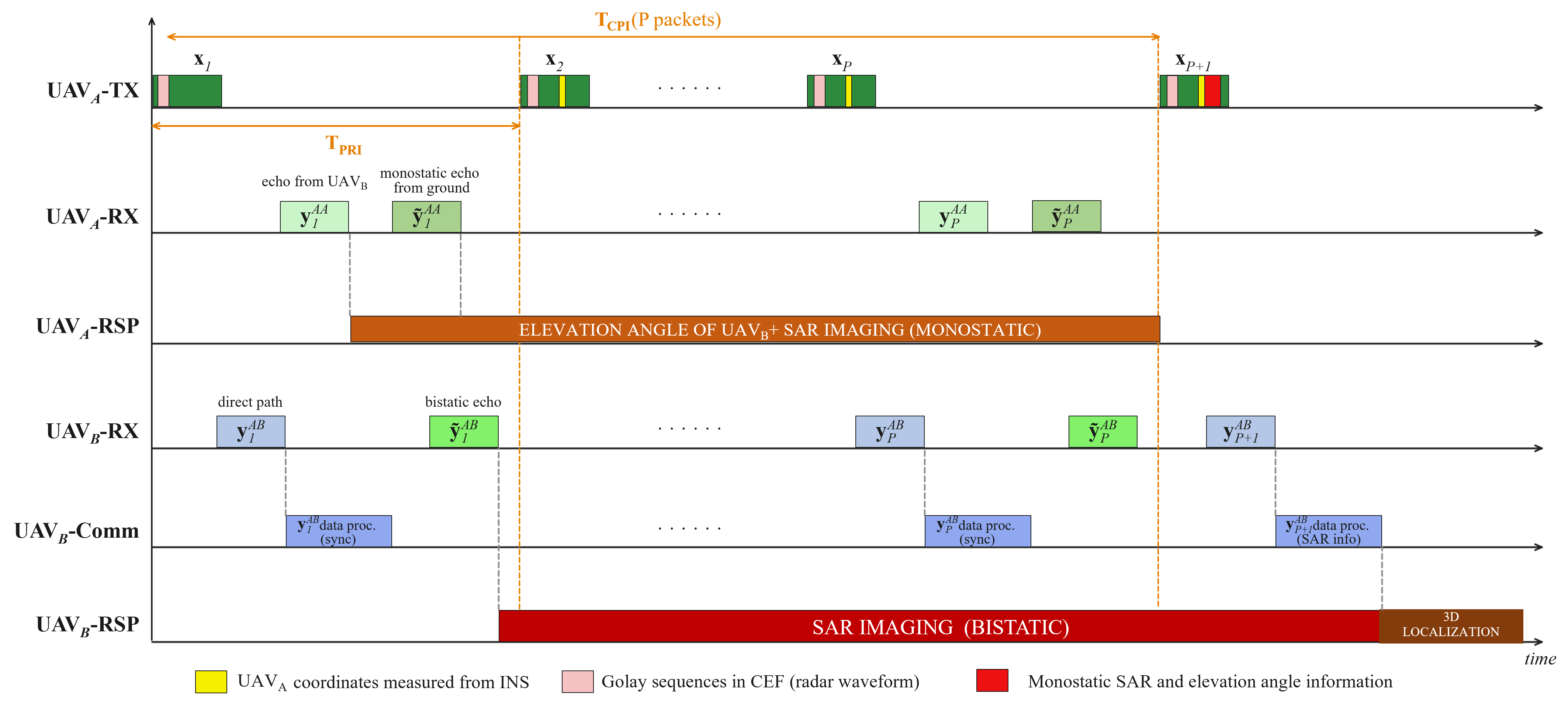}
   \vspace{-2mm}
    \caption{System architecture of bistatic UAV-ISAC framework.}
    \label{fig:system_architecture}
\end{figure*}
$UAV_A$ and $UAV_B$ observe the scene with look angles, $\theta^A$ and $\theta^B$, respectively. Under ideal operating conditions, $UAV_{A}$ and $UAV_B$ are to maintain identical uniform platform velocities along parallel paths with fixed displacement as shown in the figure. However, in practical scenarios, synchronization inaccuracies together with translational and rotational motion errors lead to deviations from the nominal trajectories, causing different velocities $\mathbf{v}^A$ and $\mathbf{v}^B$ of $UAV_{A}$ AND $UAV_B$, respectively.
The corresponding slant-range distances from $UAV_A$ and $UAV_B$ to the scene center are denoted by $\overline{R}^A$ and $\overline{R}^B$, respectively, as shown in Fig.~\ref{fig:system_model}. The bistatic angle at the target is $\beta$. The spatial separation between the two \acp{UAV} introduces angular, Doppler, and radar cross section (RCS) diversity that is exploited for bistatic tomographic reconstruction.\\
\indent The ISAC waveform chosen in this work is the IEEE 802.11 ad physical layer (PHY) packet that was originally proposed for mmWave vehicular communications \cite{ieee2009ieee}.  In prior work, a detailed description of the 802.11 ad-based ground-based monostatic ISAC system architecture for automotive scenarios was presented \cite{duggal2020doppler,sneh2023ieee,sneh2023radar,tewari2024reconfigurable,mangal2025millimeter}. In this section, we adapt the ISAC framework for the bistatic aerial SAR scenario as shown in Fig.~\ref{fig:system_architecture}. First, $UAV_A$ transmits packets:  $\mathbf{x}_1,\mathbf{x}_2, \cdots \mathbf{x}_P$, in the mmWave band as shown in the top row labeled $UAV_A$-TX in Fig.~\ref{fig:system_architecture}. Each packet spans one pulse repetition interval $T_{{PRI}}$ and the time span of $P$ packets forms a coherent processing interval, $T_{{CPI}}$. 
The single transmission signal serves simultaneously as a communication signal from $UAV_A$ to $UAV_B$ and as a radar-sensing waveform of ground-based targets. 
The master $UAV_A$ is equipped with two sets of directional receive antennas and an inertial navigation system (INS) unit, each serving a distinct purpose, as depicted in the $UAV_A$-RX row (second row) of Fig.~\ref{fig:system_architecture}. The first antenna set, directed towards $UAV_B$, is a two-element array that captures the echoes from $UAV_B$. The received signal, represented as $\mathbf{y}^{AA}_p$, is processed to determine the elevation angle of $UAV_B$ with respect to $UAV_A$, which is subsequently utilized for localization. The second directional antenna collects the backscattered radar echoes from the ground swath, which constitute the monostatic channel and are represented as $\tilde{\mathbf{y}}^{AA}_p$. The coordinates of $UAV_A$, estimated with the INS unit, are transmitted to $UAV_B$ in the subsequent packet for synchronization and localization.
Meanwhile, $UAV_B$ consists of two receiver channels, as illustrated in the $UAV_B$-RX row of Fig.~\ref{fig:system_architecture}: the first channel receives the direct-path communication signal, $\mathbf{y}^{AB}_{p}$, from $UAV_A$, while the second channel receives the bistatic reflected signal, $\tilde{\mathbf{y}}^{AB}_{p}$, from the ground-based targets. Subsequently, as shown in the $UAV_B$-Comm row of Fig.~\ref{fig:system_architecture}, the direct-path signal $\mathbf{y}^{AB}_{p}$ is processed to synchronize the two \acp{UAV}. The proposed ISAC-enabled bistatic UAV framework employs an IEEE 802.11 ad-based communication-assisted synchronization scheme to estimate and compensate for timing, frequency, and phase offsets, thereby enabling coherent bistatic operation. We present the underlying synchronization signal model and processing steps in detail in the subsequent section.\\
\indent Further, each \ac{UAV} performs local 2D SAR processing using its own received echoes as shown in the $UAV_A$ and $UAV_B$ RSP rows in Fig.~\ref{fig:system_architecture}. Specific Golay sequences within the channel estimation field (CEF) of the preamble also serve as the radar waveform for ISAC. The ordering of these sequences across multiple packets is chosen to ensure their Doppler resilience to enhance their suitability for identifying moving targets as described in \cite{duggal2019micro,duggal2020doppler}.
$UAV_A$ performs conventional two-dimensional monostatic SAR imaging in range and crossrange by processing $\mathbf{y}^{AA}_p$. $UAV_B$ processes the echoes, $\tilde{\mathbf{y}}^{AB}_{p}$, using the known bistatic geometry and the synchronization parameters obtained from the communication packet from $UAV_A$. At both platforms, MOCOMP corrects translational and rotational deviations in the \ac{UAV} trajectories, ensuring phase coherence under realistic flight dynamics. \\
\indent The range and cross-range information estimated by $UAV_A$ from the obtained 2D SAR image over the $P$ packets comprising $T_{CPI}$ is embedded in the data field of a subsequent $(P+1)^{th}$ packet (highlighted by the red data block in the first row in Fig.~\ref{fig:system_architecture}) and transmitted to $UAV_B$. This strategy avoids the challenges of raw-data transfer while enabling joint bistatic processing. The combined monostatic and bistatic observations provide sufficient spatial diversity to enable \ac{3D} localization of scatterers using our proposed novel sphere-plane intersection algorithm. We explain each of these algorithmic steps, in detail, in the following sections.
\section{Signal Model}
\label{sec:signalmodel}
In this section, we present the signal model, communication-assisted synchronization, MOCOMP, and \ac{3D} SAR processing framework for the proposed ISAC-based bistatic \ac{UAV} system. Here, $UAV_A$ transmits IEEE~802.11ad PHY packets to $UAV_B$ to synchronize the bistatic platforms, followed by bistatic SAR processing and \ac{3D} target localization over elevated terrain.

Let $\mathbf{x}$ denote the discrete-time baseband representation of the radar component of the IEEE~802.11ad PHY transmitted from $UAV_A$. Each $p^{th}$ packet is transmitted once per pulse repetition interval, $T_{PRI}$, and comprises $M$ samples corresponding to the sampling time, $T_s$, denoted as
\par\noindent\small
\begin{align}
   {\mathbf{x}_p}(t) = \sqrt{E_s}\sum_{m=0}^{M}{\mathbf{x}}[mT_s]\delta\left(t-mT_s-pT_{PRI}\right)e^{+j 2\pi f_c t}.
\end{align}
\normalsize
Here $\delta(\cdot)$ is the Dirac-delta function in the time domain $(t)$. The samples are scaled to energy $E_s$, and subsequently upconverted to the mmWave carrier frequency, $f_c$. The upconverted analog waveform is then radiated through high-gain transmit antennas towards $UAV_B$ and the ground. The radar echoes are collected at $UAV_A$ after two-way propagation to the ground swath. 
After downconversion and digitization, the received baseband signal corresponding to the $p$-th PRI from $Q$ ground-based scatterers is
 \par\noindent\small
\begin{align}
\nonumber \tilde{\mathbf{y}}_p^{AA}[m] = \sum_{q=1}^Q \alpha^{A}[q] \,
\mathbf{x}_p\!\left[m - \frac{2R^A_p[q]}{c}\right]
\exp\!\left(-j\frac{4\pi}{\lambda}R^A_p[q]\right)\\+\mathbf{\chi}^A[m] +\mathbf{\eta}^A[m].
\label{eq:rx_uav1}
\end{align} \normalsize
Here $\alpha^A[q]$ is the strength of the returns from each $q^{th}$ target incorporating the RCS and two-way path loss factor in the monostatic channel; $R^A_p[q]$ denotes the instantaneous slant-range distance between $UAV_A$ and the $q^{th}$ scatterer at slow-time index $p$; $c$ and $\lambda$ are the speed of light and the carrier wavelength, respectively; and $\mathbf{\chi}^A$ and $\mathbf{\eta}^A$ denote ground clutter present in the environment and the complex additive white Gaussian noise at the receiver of $UAV_A$, respectively. Similarly, radar echoes from $UAV_B$ are processed to estimate the range $R_p^{AB}$ and the azimuth and elevation of $UAV_B$ with respect to $UAV_A$ using standard range and angle processing with a three-element array \cite{1631807}. 

The signals transmitted by $UAV_A$ are received at $UAV_B$ through two distinct propagation mechanisms through separate antennas and receiver channels. The first component corresponds to the direct line-of-sight (LoS) one-way path from $UAV_A$ to $UAV_B$. After downconversion and digitization at the receiver, this signal is 
\begin{align}
\nonumber\mathbf{y}^{AB}_p[m] = \zeta \,
\mathbf{x}_p\!\left[m - \frac{R^{AB}_p}{c}-\Delta m\right]
\\\exp\!\left(-j\left(\frac{2\pi(f+\Delta f)}{c}R^{AB}_p+\Delta\phi\right)\right)+\mathbf{\eta}^B[m],
\label{eq:sync_error}
\end{align}

where $\zeta$ is the one-way path loss factor between the two \acp{UAV} and $\mathbf{\eta}^B$ models the complex additive white Gaussian noise (AWGN) at the receiver of $UAV_B$. Since both UAVs operate with independent local oscillators and clocks, the signal is characterized by residual timing, $\Delta m$, frequency, $\Delta f$, and phase, $\Delta \phi$, offsets. 

The second component of the received signal arises from two-way propagation of the transmitted signal to the ground, and reflected towards $UAV_B$ by $Q$ ground scatterers and gathered by the second antenna on $UAV_B$ oriented towards the ground. The bistatic received signal is 
\begin{align}
\nonumber\tilde{\mathbf{y}}^{AB}_p[m] = \sum_{q=1}^Q\alpha^B[q] \,
\mathbf{x}_p\!\left[m - \frac{R^A_p[q] + R^B_p[q]-\Delta m}{c}\right]\\
\nonumber\exp\!\left(-j\left(\frac{2\pi (f+\Delta f)}{c}(R^A_p[q] + R^B_p[q])+\Delta \phi\right)\right)+\\
\mathbf{\chi}^B[m] 
+\mathbf{\eta}^B[m].
\label{eq:rx2_uav2}
\end{align}
where $\alpha^B[q]$ denotes the complex scattering coefficient of the $q^{th}$ target, incorporating both the RCS and the bistatic path loss factor; $R^B_p[q]$ is the instantaneous slant-range distance between $UAV_B$ and the $q^{th}$ target while $\mathbf{\chi}^B$ and $\mathbf{\eta}^B$ denote ground clutter present in the environment and the complex additive white Gaussian noise at the receiver of $UAV_B$, respectively. Subsequently, both signals are forwarded to the radar signal processing block, where they are jointly used for \ac{3D} localization, as described in the later section. 
\begin{figure}[htbp]
    \centering
    \includegraphics[scale=0.5]{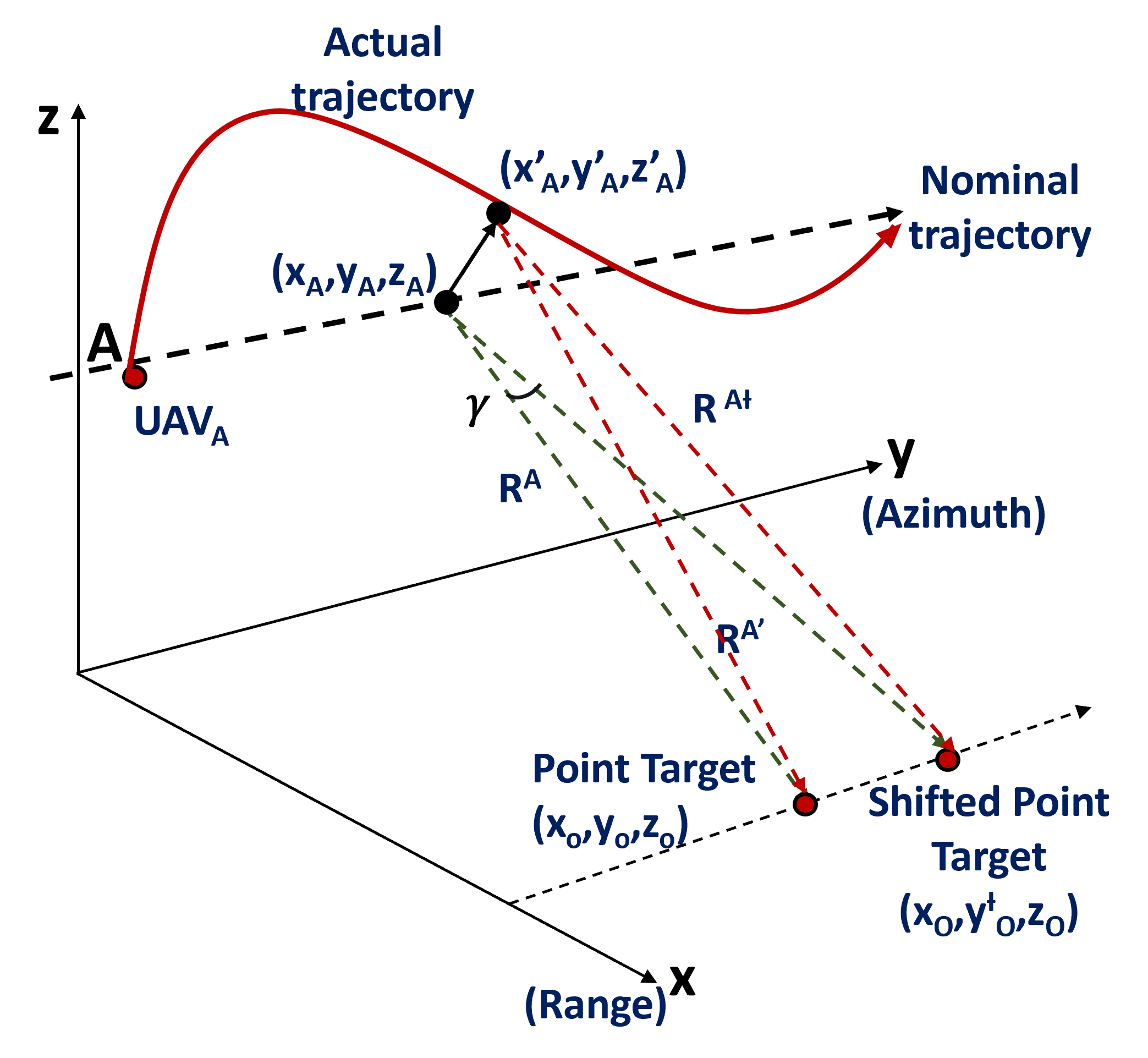}
   \vspace{-2mm}
    \caption{Incorporated translational and rotational motion errors in the proposed system model.}
    \label{fig:motion_error}
\end{figure} 
For a stationary point target located at $[x_O, y_O, z_O]$, the nominal slant range for each $p^{th}$ slow time index, $R^A[p]$, deviates from the true slant range to $R^{A\dagger}[p]$ due to translational and rotational motion errors.
The translational displacement introduces range-dependent phase errors in the received signal, resulting in slant range $R^{A'}[p]$ defocusing along the range dimension. Rotational motion, particularly yaw and pitch, causes angular/cross-range misalignment in the swath as shown in Fig.~\ref{fig:motion_error}.
Both motion errors introduce 2D geometric distortions and phase perturbations that severely impact SAR image quality and subsequent \ac{3D} localization. The combined effect of these errors can be represented through a shifted slant-range expression, as 
$R^{A\dagger}[p] = \sqrt{(R^{A'}[p])^2 + (y_O - y^\dagger_O[p])^2}$,
where the cross-range displacement $y_O - y^\dagger_O[p] = \tan\gamma[p] R^{A}[p]$ is related to the bistatic look geometry .

For clarity, the motion-error model is illustrated for $UAV_A$ and an analogous model is considered for $UAV_B$ to account for its independent translational and rotational motion errors.
Finally, the received bistatic signal modeled by jointly incorporating the effects of synchronization and UAV motion errors is 
\begin{align}
\nonumber\tilde{\mathbf{y}}^{AB}_p[m] = \sum_{q=1}^Q\alpha^B[q] \,
\mathbf{x}_p\!\left[m - \frac{R^{A\dagger}_p[q] + R^{B\dagger}_p[q]-\Delta m}{c}\right]\\
\nonumber\exp\!\left(-j\left(\frac{2\pi (f+\Delta f)}{c}\big(R^{A\dagger}_p[q] + R^{B\dagger}_p[q]\big)+\Delta \phi\right)\right)+\\
\mathbf{\chi}^B[m] 
+\mathbf{\eta}^B[m].
\label{eq:syc_err_motion_err_rx2}
\end{align}
The combined effect of all errors therefore modifies both the effective bistatic delay and the phase history of the received signal, necessitating subsequent synchronization and MOCOMP.
\section{Proposed Communication-Assisted Synchronization and Motion Compensation}
In this work, the communication link between the two \acp{UAV} is exploited to estimate and compensate the synchronization and motion errors in bistatic SAR processing. By sequentially compensating the timing, frequency, and phase offsets, both the direct and target-reflected signals received at $UAV_B$ are aligned with the transmit signal by $UAV_{A}$. Since the synchronization information is extracted from the same IEEE~802.11ad waveform used for radar sensing, no additional synchronization waveform or dedicated synchronization link is required for $UAV_B$.
\subsection{ISAC Based Synchronization}
Both the clock offset, $\Delta m$, and the exact distance between the \acp{UAV}, $R^{AB\dagger}_p$, are unknown at the receiver due to the motion errors. Since $UAV_B$ is a passive receiver, the standard two-way transfer of time information cannot be used for time synchronization.
Instead, we propose a radar-assisted, one-way passive synchronization framework that leverages the monostatic ISAC capability of $UAV_A$. In this scheme, $UAV_A$ transmits a sequence of 802.11ad 
directional frames at local timestamps $t^A_p$, which are passively received and timestamped by $UAV_B$ as $t^B_p = t^A_p + \frac{R^{AB\dagger}_p}{c} + \Delta m$, where $\frac{R^{AB\dagger}_p}{c}$ is the line-of-sight one-way propagation delay. Concurrently, $UAV_A$ 
captures the backscattered echo reflected off of $UAV_B$, estimating the round-trip motion-corrupted propagation delay corresponding to
$2R^{AB\dagger}_p/c$. Crucial to this precise estimation 
is the ultra-wideband architecture of the IEEE 802.11ad standard, which utilizes 
a high baseband sampling rate of 1.76~GHz that yields a discrete time-domain resolution of approximately 
0.57~ns based on cross-correlation of the received signal with the transmit reference signal. By feeding forward the telemetry tuple 
$\{t^A_p, R^{AB\dagger}_p\}$ within the data payload of the subsequent $(p+1)^{th}$ frame, $UAV_B$ retrospectively isolates $\Delta m$. Since the 
radar range estimation and the communication time-of-arrival acquisition 
are executed by the same physical wavefront, aerodynamic errors along translation and rotation are inherently captured and canceled,  ensuring sub-nanosecond synchronization bounds.

After compensating the timing offset, each received $p^{th}$ packet is correlated with the previous $(p-1)^{th}$ packet to obtain a final complex correlation quantity as shown
$g_p =
\sum_{m=1}^{M}
\tilde{\mathbf{y}}^{AB}_{p}[m]
\left(\tilde{\mathbf{y}}^{AB}_{p-1}[m]\right)^{*}$,
where $T_s$ is the sampling interval.  The phase of the correlation output provides an estimate of the residual carrier frequency offset (CFO) which is estimated as 
$\Delta f =\frac{1}{2\pi MT_s}\angle g_p$, where $\angle(\cdot)$ is the phase of the complex quantity. 
The estimated CFO is then compensated by applying the corresponding conjugate frequency-rotation term to the received samples, thereby removing the oscillator-induced frequency mismatch. 
Again, the waveform's high sampling rate enables fine temporal resolution, supporting accurate synchronization between the two UAVs. 

After timing and CFO compensation, the residual phase offset $\Delta \phi$, remains to be compensated. The residual phase offset is estimated from the phase of the complex correlation of the received pilot from the direct path from $UAV_A$ with the known reference as
\begin{equation}
{\Delta\phi}
=
\angle\left({\mathbf{y}}^{AB}_p[m_{pilot}]\mathbf{x}^{*}[m_{pilot}]\right),
\label{eq:phase_offset_est}
\end{equation}
where, $[m_{pilot}]$ represents the sample index corresponding to the pilot symbol.
The estimated phase offset is then compensated by applying the conjugate phase rotation to the received pilot, yielding the synchronized bistatic received signal.
\subsection{ISAC Based Motion Compensation}
After compensating for timing, frequency, and phase offsets, the synchronized bistatic signal is used for motion-error compensation. In the proposed framework, MOCOMP is performed in two consecutive stages: coarse MOCOMP followed by fine MOCOMP. The former corrects the dominant geometric phase errors, whereas the latter removes the residual phase error from the focused SAR image. Coarse MOCOMP is applied to the synchronized raw radar data before SAR processing. Onboard INS measurements on $UAV_A$ determine its deviation from the nominal trajectory. For a stationary point scatterer, the corresponding range deviation at slow-time index $p$ for $UAV_A$ is expressed as
$\Delta R^A_p=R^{A\dagger}_p-R^A_p$. 
The correction is applied to the synchronized received signal as
\begin{equation}\tilde{\mathbf{y}}^{A}_p[m]
=\tilde{\mathbf{y}}^{A\dagger}_p[m]
\exp\!\left(j\frac{4\pi}{\lambda}\Delta R^A_p\right).
\label{eq:coarse_comp}
\end{equation}
This operation aligns the measured data with the nominal bistatic trajectory while removing the dominant motion-induced phase error. 
A similar operation cannot be directly applied to $UAV_B$ since this UAV is not mounted with navigational sensors. Instead, the $UAV_B$ receives the measured $R^{A\dagger}_p$ from every $(p+1)^{th}$ communication link and estimates the bistatic distance $R^{A\dagger}_p+R^{B\dagger}_p$ through bistatic radar processing, from which it retrospectively estimates $R^{B\dagger}_p$. Since the UAV's ideal trajectory is known aprior, the coarse error correction based on $R^{B\dagger}_p-R^{B}_p$ is performed every $(p+1)^{th}$ frame. 

Subsequently, residual phase errors remain after coarse compensation because of inaccuracies. Fine MOCOMP is therefore applied after preliminary SAR focusing. While many algorithms have been discussed in the literature, we use the phase gradient autofocus \cite{wahl1994phase} that effectively trades off between residual phase-estimation accuracy and computational complexity in this work.
\section{\ac{3D} SAR Processing}
Radar echoes over each CPI are collected independently at the receivers of $UAV_A$ and $UAV_B$ as shown in 
\eqref{eq:coarse_comp}. These signals are organized into two radar rectangles comprising fast-time and slow-time samples and processed to obtain corresponding monostatic and bistatic SAR images. We apply the range-Doppler algorithm on the radar data, due to its computational efficiency and suitability for stripmap UAV geometries \cite{CummingHorneSAR}. The processing consists of range compression and range-cell migration correction to correct for range-azimuth coupling, and azimuth-domain focusing. These steps are not presented in detail since they have been discussed extensively in the literature. 
The focused monostatic and bistatic SAR images are subsequently processed with a constant false-alarm-rate detector to identify the dominant target's response. The detected highest peak corresponding to the target is then used to extract the relevant range and cross-range information for the subsequent \ac{3D} localization stage.
\subsection{Proposed Framework for 3D  Localization}
The \ac{3D} localization is performed using the proposed sphere-plane intersection algorithm enabled by the ISAC bistatic SAR framework. For this purpose, the range and cross-range information obtained from the focused monostatic 2D SAR image generated at $UAV_A$ is transmitted to $UAV_B$ through the $(P+1)^{\text{th}}$ packet, enabling $UAV_B$ to jointly process the monostatic and bistatic SAR information for subsequent \ac{3D} localization.

This localization requires the actual coordinates of $UAV_A$ and $UAV_B$, including their respective position errors. The actual coordinates of $UAV_A$ are obtained from its onboard INS and transmitted to $UAV_B$ through the ISAC communication link. Since $UAV_B$ is not equipped with an INS unit, its coordinates must instead be obtained from the available SAR and elevation information: the $y_B$ coordinate is determined from the SAR-derived monostatic and bistatic information, while the $x_B, z_B$ coordinate is obtained from the angle estimation at $UAV_A$. 

\begin{figure}[htbp]
    \centering
    \includegraphics[scale=0.25]{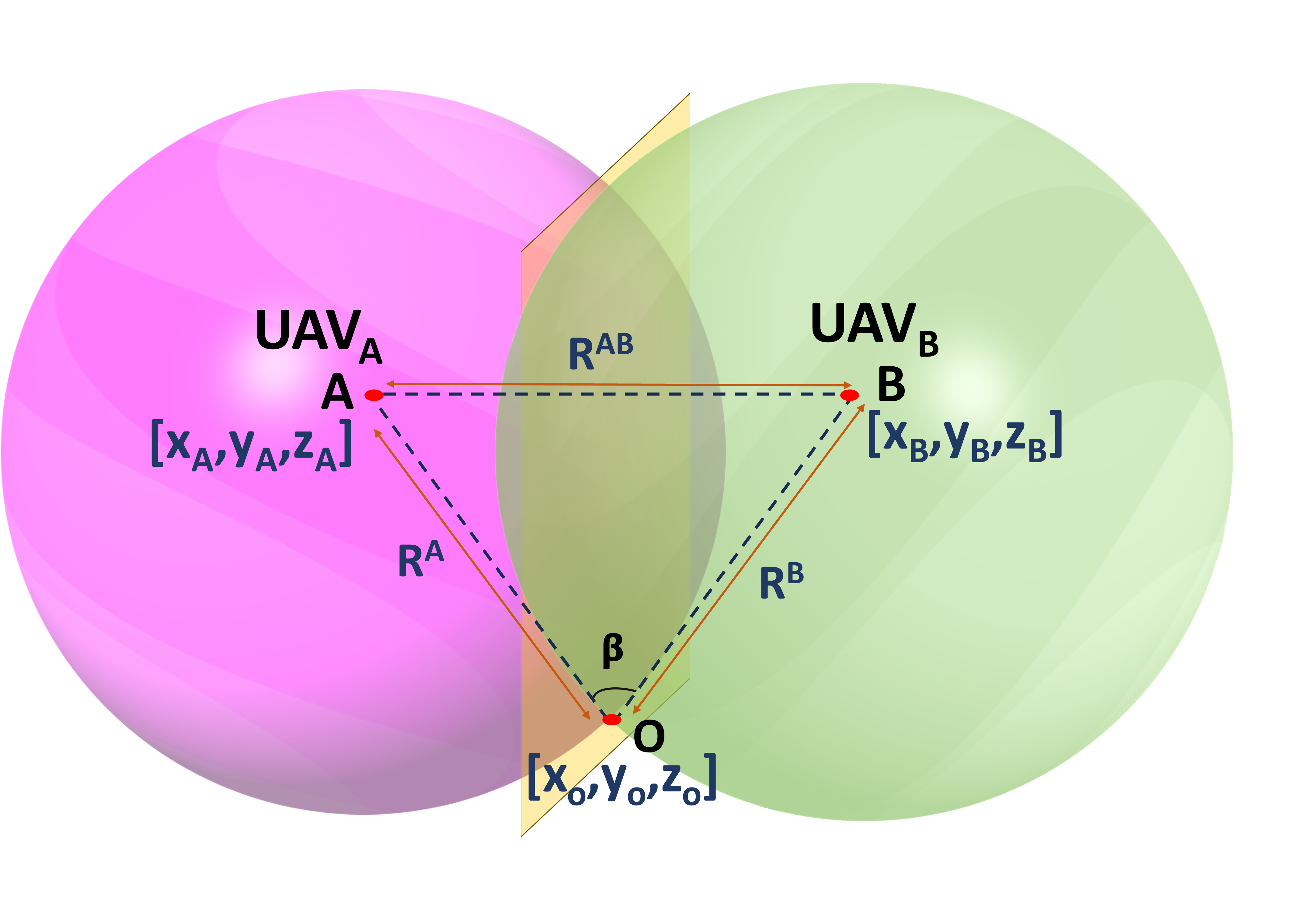}
   \vspace{-2mm}
    \caption{\ac{3D} localization based on sphere-plane intersection algorithm.}
    \label{fig:sphere_plane}
\end{figure}

Further, for ease of visualization and to facilitate the understanding of the proposed localization geometry, consider two spheres in the Cartesian coordinate system, each centered at the respective UAV positions $\mathbf{A} = [x_A, y_A, z_A]$ and $\mathbf{B}= [x_B, y_B, z_B]$, respectively, as illustrated in Fig.~\ref{fig:sphere_plane}. 
The radii of these spheres are  $R^A$ and $R^B$, respectively, and are each estimated from the slant range to the target for the monostatic and bistatic SAR images.  Simultaneously, $\mathbf{A}$, $\mathbf{B}$ and $\mathbf{O}$ also form a triangle with $R^A$ and $R^B$ as two sides of the triangle. 
The bistatic angle $\beta$ at  $\mathbf{O}$ is determined as
\begin{align}
\nonumber\beta =
\sin^{-1}\!\left[
\frac{2R^A(R^A+R^B)-(R^A+R^B)^2+R^{{AB}^2}}
{2R^AR^{AB}}
\right]
-\\
\sin^{-1}\!\left[
\frac{(R^A+R^B)^2-R^{{AB}^2}-2R^B(R^A+R^B)}
{2R^BR^{AB}}
\right].
\label{eq:beta_calc}
\end{align}
The estimated $\beta$ is then used as an angular constraint
on the unknown target position $\mathbf{O}=[x_O,y_O,z_O]$, based on Cosine law as shown in
\begin{equation}
\cos\beta =
\frac{(\mathbf{A}-\mathbf{O})\cdot(\mathbf{B}-\mathbf{O})}
{R^AR^B}.
\label{eq:cos_beta}
\end{equation}

The measured slant ranges $R^A$ and $R^B$ constrain the unknown target position $\mathbf{O}$ to lie on two spheres centered at the respective UAV positions, yielding
\begin{equation}
\|\mathbf{O}-\mathbf{A}\|^2=R^{A^2},
\qquad
\|\mathbf{O}-\mathbf{B}\|^2=R^{B^2}.
\label{eq:sphere_constraints}
\end{equation}
Subtracting the two sphere equations eliminates the quadratic unknowns and yields the plane containing their intersection circle,
\begin{align}
\nonumber 2(x_B-x_A)x_O
+2(y_B-y_A)y_O
+2(z_B-z_A)z_O
=\\
R^{A^2}-R^{B^2}
+x_B^2-x_A^2
+y_B^2-y_A^2
+z_B^2-z_A^2.
\label{eq:plane}
\end{align}
Since the cross-range coordinate $y_O$ is obtained from the focused
SAR image, the plane equation in \eqref{eq:plane} and the angular
constraint in \eqref{eq:cos_beta}, provide two equations in the two unknown
coordinates $x_O$ and $z_O$. Solving these equations yields the candidate target positions. The physically valid solution is then selected based on the known \ac{UAV}-target geometry and scene visibility
constraints.

\section{Simulation Setup}
\label{sec:simsetup}
This section describes the simulation environment used to evaluate the proposed ISAC-based bistatic UAV framework, including the radar and platform configuration, the motion errors, and the target and terrain characteristics.

\subsection{Radar and Platform Parameters}
We consider stripmap-mode operation for the proposed bistatic SAR system, with both \acp{UAV} operating at a carrier frequency of 35~GHz, chosen because it trades off between sensing resolution and propagation/link-budget requirements and is hence commonly used with mmWave UAV sensing platforms \cite{wang2005enhanced}. The PRI is $2~\mu$s, corresponding to the duration of a single IEEE 802.11 ad packet. The system provides a range resolution of 0.08~m and a cross-range resolution of 0.3~m over a ground swath of $100~\text{m} \times 100~\text{m}$. Both \acp{UAV} move with a uniform platform velocity of 50~m/s along linear trajectories parallel to the azimuth ($y$) direction, with squint angles of $0^\circ$.

We evaluate the proposed framework using Monte Carlo simulations by randomly varying the platform parameters in each realization. We vary the UAVs' look angles between $0^\circ$ and $45^\circ$, while the slant range between the \acp{UAV} and the scene center is randomly selected from 800~m to 1200~m. We benchmark the ISAC-based waveform against the conventional FMCW waveform, with both waveforms having identical radar range and cross-range resolutions, carrier frequency, effective bandwidth, pulse duration, PRI, and CPI.

To account for realistic platform motion, we introduce independent random translational velocity errors into the trajectories of both UAVs. Similarly, the velocity errors are modeled as Gaussian perturbations, with their standard deviation varied over logarithmically spaced values of ${0.5,1,2,4,8}$~m/s. These perturbations are introduced to simulate deviations from the nominal UAV velocity and to evaluate the robustness of the proposed MOCOMP implementation under varied platform motion errors.

\subsection{Target and Terrain Model}
\begin{figure}[htbp]
    \centering
    \includegraphics[scale=0.35
]{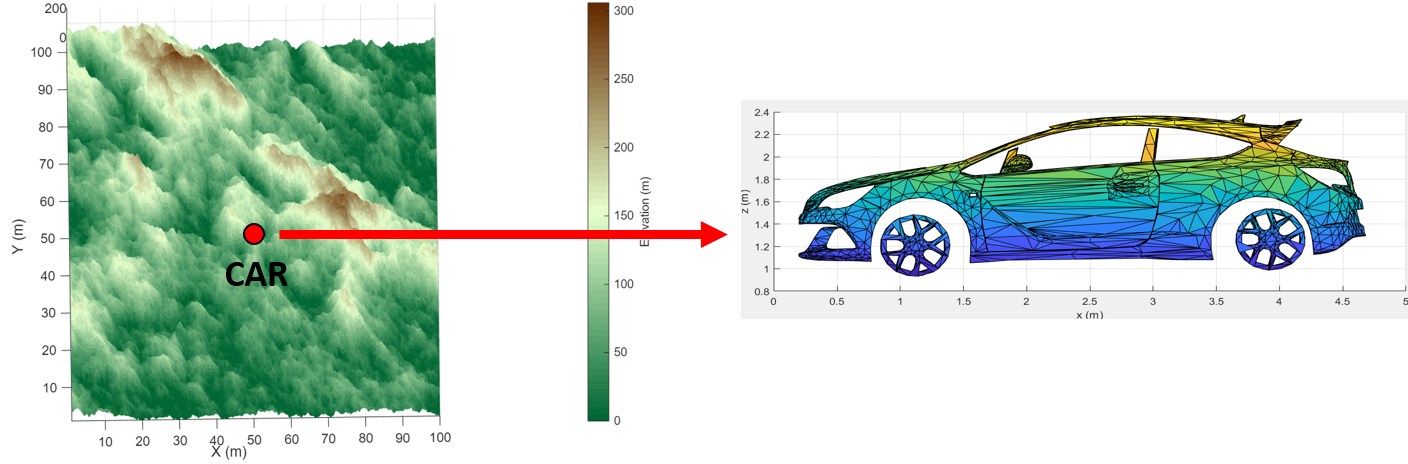}
   \vspace{-2mm}
    \caption{Automotive target (mid-size car) embedded in a digital elevation terrain model. }
    \label{fig:target_model}
\end{figure}

The target considered in this work is an extended target modeled as a static \ac{3D} automotive mid-size car of dimensions 4.5 $m$ $\times$  1.6 $m$ $\times$ 1.5 $m$ as shown in Fig.~\ref{fig:target_model}. 
The chassis, wheels, and other vehicular components are modeled using a metallic triangular mesh as described in \cite{pandey2020database,pandey2022classification}.
The centroid of each triangular facet is considered an individual scattering element, with its radar cross-section determined by the facet area and its aspect with respect to the bistatic radar angle. The complete vehicle consists of 1801 scattering elements. The total time-varying radar cross-section of the target is obtained from the superposition of the returns from all facets at each slow time instant. A binomial statistical variable is introduced as a weight for each facet to account for shadowing of parts of the vehicle from each UAV. The target's centroid position is uniformly distributed within the swath area in each Monte Carlo simulation. Similarly, its orientation is also varied. 
In the bistatic configuration, the RCS of the target is governed by the bistatic geometry and consequently depends on the distinct transmitter, $UAV_A$, and receiver, $UAV_A$, aspect angles of the target \cite{willis2005bistatic}.

A realistic uneven ground terrain spanning the simulated swath is generated with a DEM \cite{mathworks_simulated_land_sar}. The terrain elevation varies from 0~m to 20~m above sea level, across the swath, representing non-flat topography along the imaging scene as shown in Fig.~\ref{fig:target_model}. The uneven terrain characterized by the DEM gives rise to spatially distributed scattering responses, with the mean RCS varying from 0.5~${m}^2$ for low-elevation terrain to 5~${m}^2$ for peak terrain. For each DEM element, the RCS is randomly generated according to a Weibull distribution parameterized by the corresponding mean RCS. We place the automotive target over the elevated terrain to evaluate the proposed localization algorithm's ability to accurately recover the \ac{3D} target coordinates under non-planar terrain conditions with ground-based clutter. The receiver noise is modeled as additive thermal noise with a mean noise
power of $KT_0BF$, where $T_0=290~\mathrm{K}$ is the reference temperature,
$B=1.76~\mathrm{GHz}$ is the receiver bandwidth, and $F$ is the linear noise factor corresponding to a receiver noise figure of $1.76~\mathrm{dB}$. 
\section{Performance Analysis}
\label{sec:results}
\subsection{Timing Analysis of the Proposed Architecture}
Table~\ref {tab:timing_analysis} depicts the timing breakdown of the proposed ISAC-based bistatic UAV framework. Here, we analyze the end-to-end latency of communication-assisted synchronization, INS-based coordinate acquisition, elevation-angle estimation, MOCOMP, SAR image formation, and \ac{3D} localization. The framework's total latency depends on the number of packets, their transmission duration, and the per-stage processing time. All processing is carried out in floating-point in MATLAB 2025a on an Intel Core i7 processor with 128~GB RAM.

Both \acp{UAV} move along the positive azimuth dimension with a total flight path length of $100$~m at a constant velocity of $50$~m/s. With a PRI of $0.66$~ms, a total of $3026$ packets are transmitted within a CPI, of which the first $P = 3025$ packets carry newly sensed radar and communication data, while the final $(P+1)^{th}$ packet, $\mathbf{x}_{P+1}$, carries the processed information accumulated over the CPI. The corresponding timing breakdown for each processing stage is summarized in Table~\ref{tab:timing_analysis}.
Each of the $3026$ packets received by $UAV_B$ is processed to extract timing, carrier-frequency, and phase-offset information for communication-based synchronization, which takes $1.6$~ms per packet and is executed in parallel upon packet reception, thereby avoiding cumulative delay across the CPI. At $UAV_A$, two additional per-packet operations are carried out over the $3025$ packets: the navigational unit measures $UAV_A$'s own coordinates, requiring $0.02$~ms per packet, and this information is embedded in the data field of the consecutive transmitted packet. On the other hand, echo received from $UAV_B$ is processed to estimate the elevation angle between the two \acp{UAV}, requiring $0.2$~ms per packet and the resulting $P$ elevation-angle estimates are accumulated and embedded, together with the monostatic SAR information, into the data field of the $(P+1)^{th}$ packet for transmission to $UAV_B$.

Note that communication-based synchronization is not required at the $UAV_A$ receiver; the received echoes are sent directly to the radar signal processing block after digitization and downconversion. The coarse MOCOMP applied to the raw data takes $1.1$~ms per packet for both the monostatic and bistatic channels. In the monostatic case, range compression requires $1.2$~ms per pulse and is processed in parallel, since the pulses are independent and thus does not introduce serial latency accumulation. In the bistatic case, range compression is performed together with range-offset compensation, requiring $1.4$~ms per pulse, also processed in parallel. After all $P$ packets within the CPI are collected, azimuth compression across slow-time samples and fine MOCOMP are each performed once, common to both channels, requiring $1.2$~ms and $1.2$~ms, respectively ($2.4$~ms in total).
After 2D SAR processing, $UAV_B$ performs data fusion upon receiving the accumulated elevation-angle, actual coordinates of $UAV_A$, and monostatic SAR information in the $(P+1)^{th}$ packet, and subsequently performs \ac{3D} localization using the proposed sphere-plane intersection algorithm, which requires $1.1$~ms. For comparison, a conventional InSAR-based height-estimation approach requires $2.3$~ms for the same stage, all other processing stages being identical between the two methods.

Since communication-based synchronization, INS measurement, elevation-angle estimation, coarse MOCOMP, and range compression are all executed in parallel with ongoing packet transmission, the dominant contributor to the total latency is the CPI duration itself, $3026 \times 0.66~\text{ms} \approx 1997.16$~ms. The overall end-to-end latency thus adds up to $2006.18$~ms for the proposed method and $ 2007.38$~ms for the InSAR-based localization, respectively.

\begin{table}[htbp]
\centering
\caption{Timing breakdown of the proposed ISAC-based bistatic framework for a flight path duration of 2s of both UAVs.}
\label{tab:timing_analysis}
\begin{tabularx}{\columnwidth}{@{} >{\raggedright\arraybackslash}X c c c @{}}
\toprule
\textbf{Signal /} & \textbf{Number of} & \multicolumn{2}{c}{\textbf{Duration (ms)}} \\ 
\cmidrule(lr){3-4} 
\textbf{Processing Stage} & \textbf{Packets} & \textbf{Proposed method} & \textbf{InSAR} \\ 
\midrule
IEEE~802.11ad packet transmission (PRI) & 3026 & 0.66 each & 0.66 each \\ 
INS measurements of $UAV_A$ coordinates  (at $UAV_A$) & 3025 & 0.02 each$^{\ast}$ & 0.02 each$^{\ast}$ \\
Communication based synchronization (at $UAV_B$) & 3026 & 1.6 each$^{\ast}$ & 1.6 each$^{\ast}$ \\ 
Elevation angle estimation of $UAV_B$   (at $UAV_A$) & 3025 & 0.2 each$^{\ast}$ & 0.2 each$^{\ast}$ \\ 
Coarse MOCOMP (Monostatic/Bistatic) & 3025 & 1.1 each$^{\ast}$ & 1.1 each$^{\ast}$ \\ 
Range compression (Monostatic) & 3025 & 1.2 each$^{\ast}$ & 1.2 each$^{\ast}$ \\ 
Range compression (Bistatic) & 3025 & 1.4 each$^{\ast}$ & 1.4 each$^{\ast}$ \\ 
Azimuth compression (Monostatic/Bistatic) & 1 & 1.2 & 1.2 \\ 
Fine MOCOMP (Monostatic/Bistatic) & 1 & 1.2 & 1.2 \\ 
\ac{3D} localization & 1 & 1.1 & 2.3 \\ 
\midrule
\textbf{Total End-to-End Latency} & -- & \textbf{2006.18} & \textbf{2007.38} \\ 
\bottomrule
\multicolumn{4}{@{}l@{}}{$^{\ast}$\footnotesize Parallel processing applied per packet.}
\end{tabularx}
\end{table}

\subsection{Two-Dimensional Localization Performance Analysis}
\begin{figure*}[htbp]
    \centering
    \includegraphics[scale = 0.5]{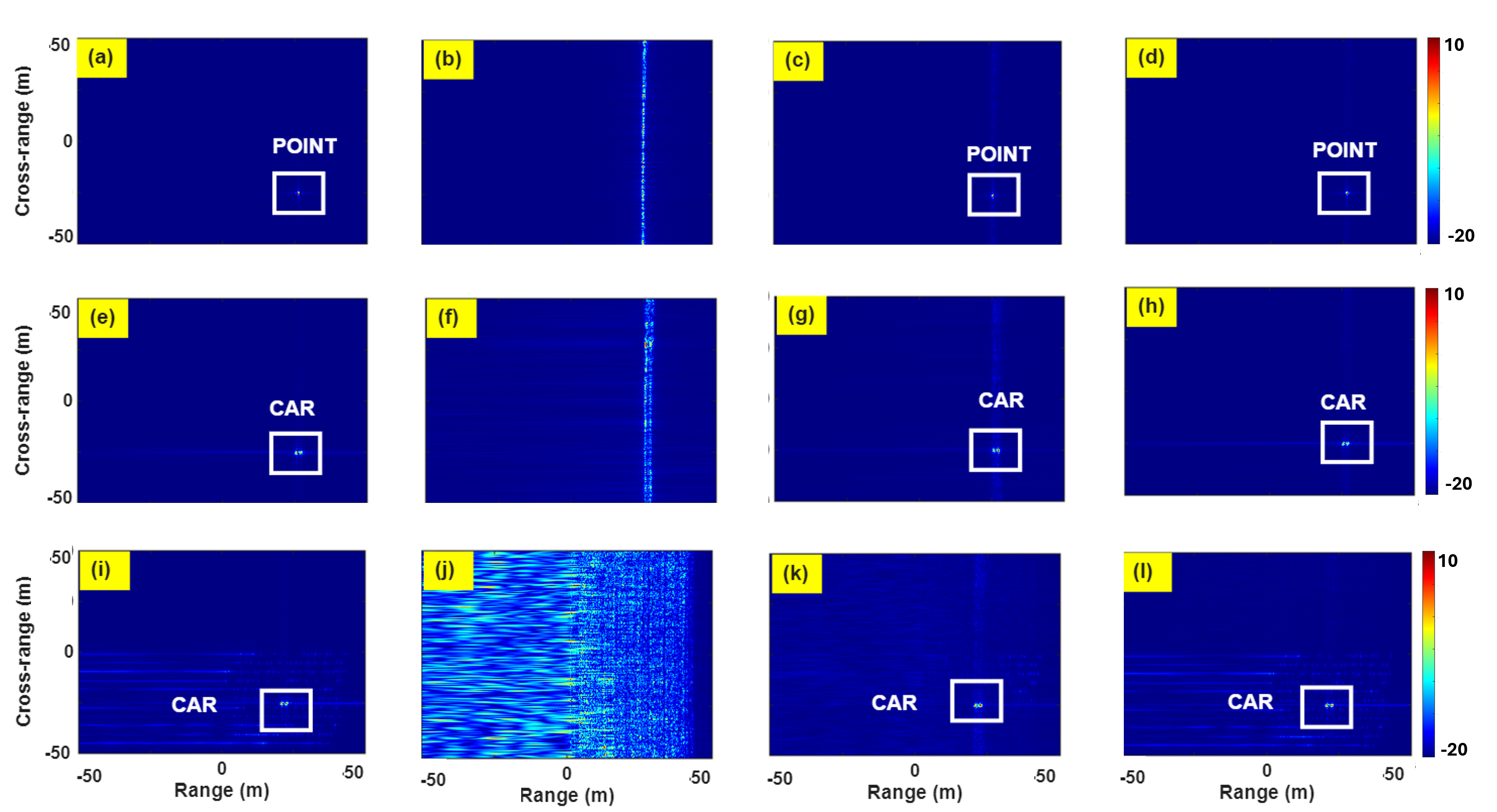}
  \caption{SAR image formation results illustrating the impact of motion errors and MOCOMP. (a)–(d) Point scatterer: (a) ideal case without motion errors, (b) with translational and rotational motion errors, (c) after coarse MOCOMP, and (d) after fine MOCOMP. (e)–(h) Mid-size car located on flat terrain (zero elevation) under the same four processing conditions. (i)–(l) Automotive target placed on elevated terrain, where image formation accounts for terrain variations using a digital elevation model. The white boxes highlight the target regions.}
  \label{fig:2D_results}
\end{figure*}
A qualitative and quantitative evaluation of the 2D SAR imaging performance for the proposed bistatic \acp{UAV} framework is presented in this section. Fig.~\ref{fig:2D_results} illustrates the focused SAR images for the three different targets considered: (i) a point scatterer on a flat ground, (ii) a mid-size automotive target placed on flat terrain, and (iii) the same automotive target positioned on an uneven terrain modeled using a DEM. For each scenario, four results are shown: ground truth images, images where the simulation model is incorporated with translational and rotational motion errors, images after coarse MOCOMP, and images after fine MOCOMP. 

Figs.~\ref{fig:2D_results}(a)-(d) present the 2D SAR viewgraphs of a single point scatterer for qualitative comparison. In the ideal scenario shown in Fig.~\ref{fig:2D_results}(a), the target is accurately focused at the range and cross-range location corresponding to the ground truth values. In the presence of translational and rotational motion errors, significant defocusing is observed in Fig.~\ref{fig:2D_results}(b). The energy spreads along the azimuth dimension due to uncompensated phase errors, resulting in peak broadening and sidelobe distortion.
Fig.~\ref{fig:2D_results}(c) shows the viewgraph after applying coarse MOCOMP where the major range migration induced by translational errors is largely compensated, restoring approximate target localization. However, residual phase errors remain due to higher-order motion components and rotational deviations. These errors are effectively mitigated by fine MOCOMP based on phase gradient autofocus as shown in Fig.~\ref{fig:2D_results}(d), resulting in a well-focused point target comparable to the ideal case.
Figs.~\ref{fig:2D_results}(e)-(h) show the results for the mid-size car on a flat terrain (zero elevation). In the ideal case (e), the distributed scattering from the extended target is clearly visible, with prominent reflections corresponding to strong scattering components such as the roof, chassis, and wheels. The spatial extent of the vehicle is accurately represented in both range and cross-range.
When motion errors are introduced (f), the distributed scattering pattern becomes smeared and distorted. The defocusing is more pronounced than in the point-target case due to the interference between multiple scatterers. After coarse MOCOMP (g), the dominant geometric distortions are corrected, but minor blurring remains. Fine MOCOMP (h) restores image sharpness, demonstrating that the proposed two-stage MOCOMP framework effectively preserves structural fidelity for distributed automotive targets on flat terrain.

Figs.~\ref{fig:2D_results}(i)-(l) present the automotive target placed on an uneven terrain modeled using a DEM. Unlike the flat-terrain scenario, the target now possesses a non-zero elevation component. In the ideal processing case without motion errors (i), the target appears focused; however, the apparent range position is incorrect. This discrepancy arises because the slant range measured by the radar corresponds to the three-dimensional Euclidean distance between the \acp{UAV} and the elevated scatterer. When projected onto a two-dimensional ground plane under a flat-Earth assumption, the height component is coupled with the horizontal coordinates. Consequently, the target appears shifted in the range dimension relative to its true ground projection. This geometric misinterpretation persists even in the absence of motion errors.
In the presence of motion errors (j), the image is further degraded due to phase distortions, compounding both geometric and focusing inaccuracies. Coarse MOCOMP (k) corrects the dominant motion-induced range shifts but cannot compensate for the inherent geometric projection error. After fine MOCOMP (l), the target is sharply focused, yet the range misplacement remains. The results in Figs.~\ref{fig:2D_results}(i)-(l) highlight a fundamental limitation of conventional 2D SAR imaging: while MOCOMP can restore phase coherence and improve focusing quality, it cannot resolve height-induced geometric ambiguity. The 2D SAR image inherently represents reflectivity as a function of range and cross-range under a single-look geometry. When the target lies on elevated terrain, the measured slant range does not uniquely determine its horizontal ground position unless the elevation is known or estimated. Therefore, even perfectly focused 2D SAR images may exhibit incorrect ground-range localization for elevated targets. This limitation is addressed by the proposed bistatic tomographic reconstruction using the proposed \ac{3D} localization framework, which explicitly exploits multi-view geometry to separate horizontal and vertical components of the target position. Thus, while the proposed motion compensation framework effectively restores image focus under realistic \acp{UAV} dynamics, accurate localization of targets on non-planar terrain requires three-dimensional processing beyond conventional 2D SAR imaging.

Next, we quantitatively assess the focusing performance by computing the peak sidelobe ratio (PSLR) and integrated sidelobe ratio (ISLR) for each case in Table~\ref{tab:pslr_islr_results}. We observe that the ideal case without any motion error exhibits the best sidelobe suppression for all the targets, with PSLR values approaching the theoretical limits. The introduction of translational and rotational motion errors significantly degrades both PSLR and ISLR due to phase incoherence and energy spreading across range and azimuth dimensions.
Application of coarse MOCOMP substantially improves image quality by correcting dominant range migration effects; however, residual phase errors continue to limit sidelobe suppression. After fine MOCOMP, PSLR and ISLR values closely approach those of the ideal case, demonstrating the effectiveness of the two-stage MOCOMP framework.
For the elevated terrain scenario, although sidelobe metrics after fine MOCOMP are comparable to the flat-terrain case, the range mis-registration discussed earlier persists due to geometric projection ambiguity. This confirms that while MOCOMP restores image sharpness, conventional 2D SAR processing remains inherently incapable of resolving elevation-induced localization errors. 

\begin{table*}[htbp]
\centering
\caption{2D SAR quantitative analysis under different motion compensation scenarios.}
\label{tab:pslr_islr_results}
\begin{tabular}{|l|cc|cc|cc|cc|}
\hline
\multirow{2}{*}{\textbf{Target Scenario}} 
& \multicolumn{2}{c|}{\textbf{No Motion Error}} 
& \multicolumn{2}{c|}{\textbf{With Motion Error}} 
& \multicolumn{2}{c|}{\textbf{Coarse MOCOMP}} 
& \multicolumn{2}{c|}{\textbf{Fine MOCOMP}} \\ \cline{2-9}
& PSLR (dB) & ISLR (dB) 
& PSLR (dB) & ISLR (dB) 
& PSLR (dB) & ISLR (dB) 
& PSLR (dB) & ISLR (dB) \\ \hline

Point on Flat Terrain 
& -20.75 & -14.65 
& -11.7 & 7.9 

& -16.74 & -13.92 
& -19.25 & -14.58 \\ \hline

Car on Flat Terrain 
& -20.31 & -12.87 
& -6.79 & 9.70 
& -15.41 & -11.80 
& -19.10 & -11.24 \\ \hline

Car on Elevated Terrain 
& -18.52 & -10.64 
& -3.95 & 9.21 
& -12.71 & -8.55 
& -17.13 & -9.86
\\ \hline

\end{tabular}
\end{table*}
\subsection{Three-Dimensional Localization Performance Analysis}
Monte Carlo simulations are performed to analyze the \ac{3D} localization accuracy under varying platform velocity errors. For each simulation experiment, the standard deviation of the velocity is fixed, and 1000 independent realizations are considered. The experiments are repeated for 5 different velocity standard deviations: [0.5, 1, 2, 4, 8] $m/s$ . The performance is assessed for the RMSE of the estimated target coordinates $(x,y,z)$. The proposed \ac{3D} localization approach is benchmarked against a conventional interferometric SAR (InSAR) method, where two monostatic SAR images are acquired using two \acp{UAV} with a vertical baseline separation of 10~m, while all other parameters remain unchanged. The analysis is then repeated for both ISAC and FMCW waveforms, with and without MOCOMP, to investigate the combined impact of waveform and motion-error mitigation on localization performance.
\begin{figure}[htbp]
    \centering
    \includegraphics[scale=0.45]{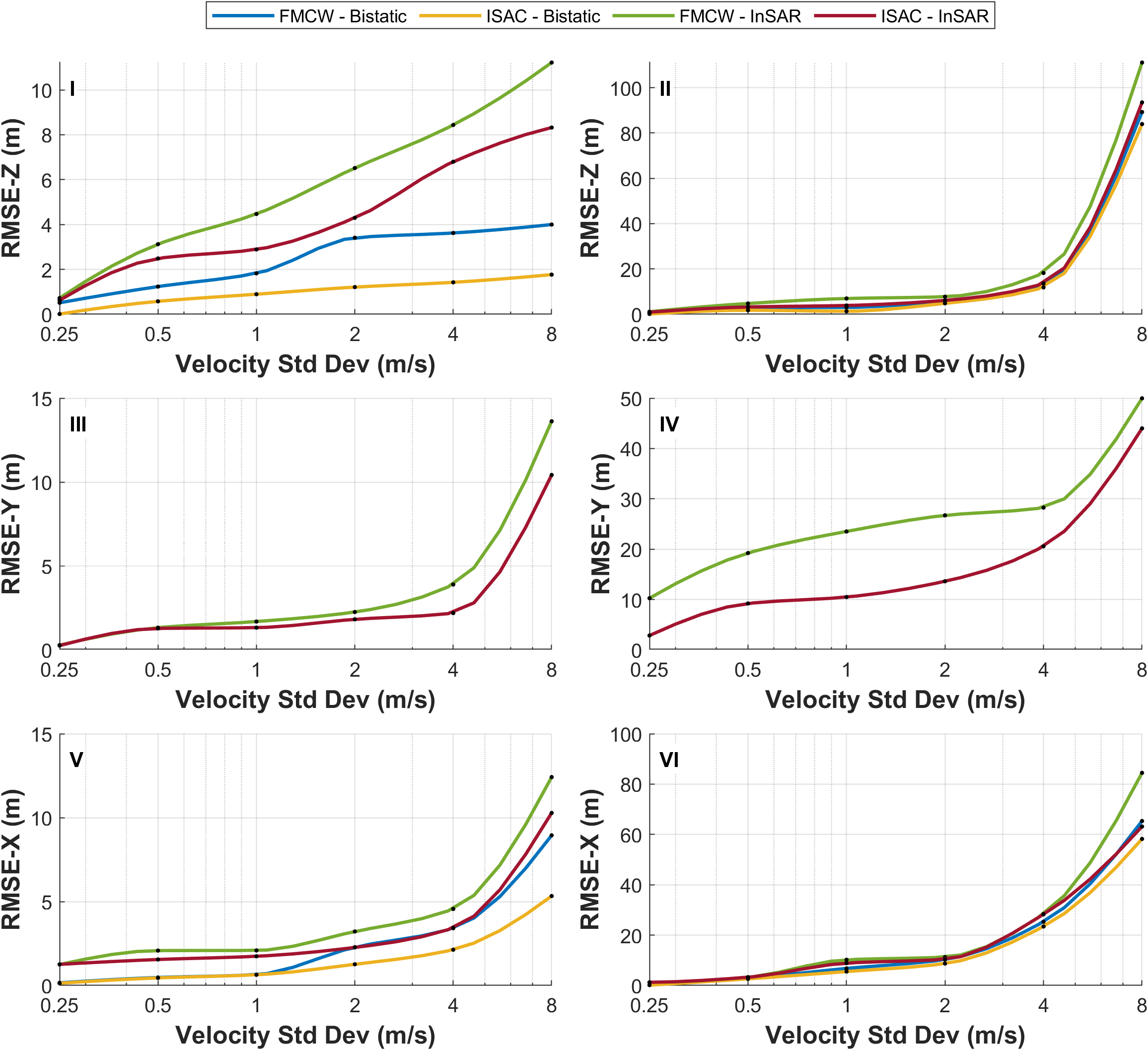}
    \vspace{-2mm}
    \caption{Radar RMSE comparison of estimated $(x,y,z)$ target coordinates for the proposed bistatic localization framework and the conventional InSAR approach under varying platform velocity errors.}
    \label{fig:3D_rmse_results}
\end{figure}

Figs.~\ref{fig:3D_rmse_results}(i), (iii), and (v) illustrate the localization RMSE corresponding to the $(x,y,z)$ coordinates when MOCOMP is incorporated. In all three viewgraphs, the results demonstrate that the proposed ISAC-enabled bistatic framework consistently achieves lower localization errors compared with the conventional InSAR baseline across all considered error levels. In particular, the improvement is most significant for estimating the elevation ($z$) coordinate, where the proposed framework maintains substantially lower RMSE values for both waveforms. This behavior arises because the proposed localization framework analytically exploits bistatic spatial diversity, thereby providing improved robustness and stability in elevation reconstruction. On the other hand, in InSAR-based localization, the elevation information is derived from interferometric phase differences, making it more sensitive to system errors. For the cross-track $y$ coordinate, the observed radar RMSE values are relatively similar between the proposed framework and the InSAR approach when identical waveforms are employed. This is primarily because both methods rely on 2D SAR image formation for azimuth estimation. In contrast, the $x$ coordinate estimates show lower error for the proposed method, since the localization of the ground-range component in the InSAR approach is strongly coupled with elevation estimation errors. We also observe that the ISAC waveform consistently achieves lower localization errors compared to the FMCW waveform under identical operating conditions due to the superior sidelobe suppression and higher PSLR performance of the Golay sequences within the ISAC waveform than the FMCW waveform \cite{sneh2023ieee}.

Figs.~\ref{fig:3D_rmse_results}(ii), (iv), and (vi)  present the localization performance in the absence of MOCOMP. In this scenario, all methods experience noticeable degradation in localization accuracy due to uncompensated translational and rotational motion errors in the UAV trajectories. These motion-induced deviations introduce phase distortions that adversely affect coherent SAR focusing and target localization. Consequently, the radar RMSE values increase significantly for all spatial coordinates compared with the motion-compensated case. Nevertheless, the proposed ISAC-based bistatic framework continues to outperform the InSAR-based approach even without MOCOMP, indicating that the proposed geometric localization strategy inherently offers greater robustness against synchronization and motion uncertainties.

Overall, the results validate that the use of over-the-air ISAC-based synchronization and MOCOMP and bistatic geometric diversity significantly enhances the robustness and accuracy of \ac{3D} target localization. The proposed framework consistently achieves lower radar RMSE values across all spatial coordinates, particularly under realistic motion-error conditions, thereby demonstrating its suitability for practical UAV-based bistatic SAR imaging applications.
\subsection{Effect of Communication Modulation Schemes on Synchronization}
In the proposed framework, the communication link is used to estimate and compensate for the timing offset, carrier-frequency offset and phase offset between the two UAVs and subsequent MOCOMP. Thus, the impact of communication-based synchronization and MOCOMP on localization performance is important to investigate. In this section, we analyze the performance across four different modulation schemes: binary phase shift keying (BPSK), quadrature phase shift keying (QPSK), 8-phase shift keying (8-PSK), and 16-quadrature amplitude modulation (16-QAM). 
In this analysis, the standard deviation of the platform velocity error is fixed at $0.5$~m/s. The RMSE values of the estimated target coordinates are computed for the proposed \ac{3D} localization method with the MOCOMP framework under each modulation scheme. Further, we have analyzed the synchronization performance in terms of bit-error rate (BER) for four different communication signal-to-noise ratio (SNR) levels: $0$~dB, $5$~dB, $10$~dB, and $15$~dB. Since the radar echo undergoes two-way propagation via the target, whereas the communication signal undergoes one-way propagation between the UAVs, the radar SNR is approximately $15$~dB lower than the communication SNR for the same noise power. Accordingly, the corresponding radar SNR levels are $-15$~dB, $-10$~dB, $-5$~dB, and $0$~dB, respectively. The results show that lower-order modulation schemes provide improved localization accuracy due to greater robustness to phase noise and synchronization errors. At a communication SNR of $15$~dB (radar SNR of $0$~dB), all modulation schemes exhibit very low  BERs of $1\times10^{-3}$, resulting in nearly identical RMSE values of approximately $0.12$~m for each coordinate. This indicates that reliable synchronization information is extracted from the communication waveform, and that accurate localization is achieved even at a radar SNR of $0$~dB. 
Further, for communication at SNR of $10$~dB (radar SNR of $-5$~dB), lower modulation schemes such as BPSK and QPSK result in a very low BER of $1.25\times10^{-3}$. Notably, the RMSE for BPSK and QPSK remains at $0.12$~m despite the $5$~dB reduction in radar SNR, indicating that the localization accuracy is governed primarily by the synchronization quality rather than the radar SNR in this regime. However, in contrast,
higher-order modulation schemes such as 8-PSK and 16-QAM exhibit noticeably higher BER values of $1.25\times10^{-2}$ and $2.75\times10^{-2}$ at the same SNR level respectively, which consequently leads to larger localization errors. Specifically, the radar RMSE increases to $1.68$~m for modulation scheme of 8-PSK and further to $12.57$~m for 16-QAM, demonstrating the sensitivity of the radar-based localization to communication reliability. 

The effect becomes more pronounced at lower communication SNR of $5$~dB, corresponding to a radar SNR of $-10$~dB. Here, BPSK maintains a relatively small BER of $7.50\times10^{-3}$ and a radar RMSE of $0.58$~m, indicating that the proposed localization framework remains stable even in degraded communications and at a radar SNR of $-10$~dB. QPSK experiences a noticeable performance degradation, with the BER increasing to $3.50\times10^{-2}$ and the corresponding radar RMSE rising to $14.54$~m. For higher-order modulation schemes, the BER increases significantly to $9.04\times10^{-2}$ for 8PSK and $1.03\times10^{-1}$ for 16-QAM respectively. At these high error rates, the synchronization information becomes unreliable, resulting in synchronization failure \textbf{(SF)} during the localization process. Consequently, accurate \ac{3D} target localization cannot be achieved under these conditions. At the lowest communication SNR of $0$~dB, which corresponds to a radar SNR of $-15$~dB, the synchronization performance degrades considerably due to the severe impact of noise on the communication link, while the radar echo itself is also significantly weakened. Under this condition, BPSK exhibits a BER of $1.08\times10^{-1}$, resulting in a significantly larger RMSE of $43.28$~m, which indicates notable degradation in localization accuracy due to the combined effect of synchronization errors and low radar SNR. For QPSK, the BER further increases to $2.03\times10^{-1}$, leading to synchronization failure and hence localization cannot be carried out. Similarly, higher-order modulation schemes such as 8PSK and 16-QAM experience very high BER values of $3.34\times10^{-1}$ and $5.01\times10^{-1}$, respectively, which disrupt the synchronization required for localization. 
Overall, these results demonstrate that reliable communication-assisted synchronization is critical to enabling accurate \ac{3D} localization in the proposed bistatic SAR framework, even when the radar SNR is as low as $0$~dB to $-10$~dB. Lower-order modulation schemes such as BPSK and QPSK provide stable synchronization even with moderate SNR, thereby enabling robust localization. In contrast, higher-order modulation schemes introduce significant synchronization uncertainty due to increased BER, which directly impacts localization accuracy. These findings demonstrate that the proposed ISAC-enabled bistatic tomographic framework effectively leverages communication signals for synchronization while maintaining reliable \ac{3D} target localization, particularly when robust modulation schemes are employed.
\begin{table*}[!t]
\centering

\caption{Performance comparison ( BER and Radar RMSE) of different modulation schemes as a function of communication metrics.}
\label{tab:modulation_RMSE}
\resizebox{\textwidth}{!}{
\begin{tabular}{|c|c|cccc|cccc|cccc|cccc|}
\hline
\multirow{2}{*}{Comm SNR} 
& \multirow{2}{*}{Radar SNR}
& \multicolumn{4}{c|}{BPSK} 
& \multicolumn{4}{c|}{QPSK} 
& \multicolumn{4}{c|}{8-PSK} 
& \multicolumn{4}{c|}{16-QAM} \\ \cline{3-18}

& 
& BER & RMSE X & RMSE Y & RMSE Z
& BER & RMSE X & RMSE Y & RMSE Z
& BER & RMSE X & RMSE Y & RMSE Z
& BER & RMSE X & RMSE Y & RMSE Z \\ \hline

15 dB & 0 dB
& $1.00\times10^{-3}$ & 0.12 & 0.12 & 0.12
& $1.00\times10^{-3}$ & 0.12 & 0.12 & 0.12
& $1.00\times10^{-3}$ & 0.12 & 0.12 & 0.12
& $1.00\times10^{-3}$ & 0.12 & 0.12 & 0.12 \\ \hline

10 dB & -5 dB
& $1.00\times10^{-3}$ & 0.12 & 0.12 & 0.12
& $1.25\times10^{-3}$ & 0.12 & 0.12 & 0.12
& $1.25\times10^{-2}$ & 1.68 & 1.68 & 1.68
& $2.75\times10^{-2}$ & 12.57 & 12.57 & 12.57 \\ \hline

5 dB & -10 dB
& $7.50\times10^{-3}$ & 0.58 & 0.58 & 0.58
& $3.50\times10^{-2}$ & 14.54 & 14.54 & 14.54
& $9.04\times10^{-2}$ & SF & SF & SF
& $1.03\times10^{-1}$ & SF & SF & SF \\ \hline

0 dB & -15 dB
& $1.08\times10^{-1}$ & 43.28 & 43.28 & 43.28
& $2.03\times10^{-1}$ & SF & SF & SF
& $3.34\times10^{-1}$ & SF & SF & SF
& $5.01\times10^{-1}$ & SF & SF & SF \\ \hline

\end{tabular}
}
\end{table*}
\section{Conclusion}
\label{sec:Conclusion}
This work presented an ISAC-enabled bistatic UAV TomoSAR framework for \ac{3D} localization of complex targets over elevated terrain. Unlike conventional approaches that employ separate communication and sensing resources or rely on dedicated synchronization infrastructure, the proposed framework utilizes the same IEEE~802.11ad waveform, RF hardware, and spectral resources for both communication and radar sensing. Communication-assisted synchronization between the distributed UAV platforms enables the estimation and compensation of timing and frequency offsets, and subsequent MOCOMP, thereby establishing the coherence required for bistatic SAR processing.
The 2D SAR results demonstrated the effectiveness of the proposed framework in recovering images in the presence of translational and rotational motion errors. Furthermore, extensive Monte Carlo simulations were conducted to evaluate the \ac{3D} localization method in terms of RMSE under different levels of platform velocity uncertainty to benchmark the proposed method against the InSAR-based approach using both ISAC and FMCW waveforms. The results demonstrated that the proposed ISAC-enabled bistatic framework provides improved localization accuracy. Finally, the interplay between radar and communication metrics was studied by analyzing the effect of communication link metrics on the radar localization performance for different modulation and coding schemes and SNR levels. 
The implementation and simulation MATLAB code developed for this work are available at \url{https://github.com/vivaswan-vm/Bistatic-UAV-3D-Sar-Imaging}

\bibliographystyle{ieeetran}
\bibliography{reference}    

\end{document}